\documentclass[aps,twocolumn,pra,superscriptaddress,showpacs,tightenlines]{revtex4}
\usepackage{amssymb}
\usepackage{amsmath}
\usepackage{graphicx}
\usepackage{epsfig}
\usepackage{subfigure}
\usepackage{amsfonts}

\begin{document}

\title{Quantum theory for spatial motion of polaritons in inhomogeneous
fields}
\author{Lan Zhou}
\affiliation{Department of Physics, Hunan Normal University, Changsha 410081, China}
\affiliation{Institute of Theoretical Physics, Chinese Academy of Sciences, Beijing,
100080,China}
\author{Jing Lu}
\affiliation{Department of Physics, Hunan Normal University, Changsha 410081, China}
\author{D. L. Zhou}
\affiliation{Institute of Physics,Chinese Academy of Sciences, Beijing 100080, China}
\author{C. P. \surname{Sun}}
\email{suncp@itp.ac.cn}
\homepage{http://www.itp.ac.cn/~suncp}
\affiliation{Institute of Theoretical Physics, Chinese Academy of Sciences, Beijing,
100080,China}

\begin{abstract}
Polaritons are the collective excitations of many atoms dressed by resonant
photons, which can be used to explain the slow light propagation with the
mechanism of electromagnetically induced transparency. As quasi-particles,
these collective excitations possess the typical feature of the matter
particles, which can be reflected and deflected by the inhomogeneous medium
in its spatial motion with some velocity. In this paper we develop a quantum
theory to systematically describe the spatial motion of polaritons in
inhomogeneous magnetic and optical fields. This theoretical approach treats
these quasi-particles through an effective Schr\"{o}dinger equation with
anisotropic depression that the longitudinal motion is like a
ultra-relativistic motion of a ``slow light velocity'' while the transverse
motion is of non-relativity with certain effective mass. We find that, after
passing through the EIT medium, the light ray bends due to the
spatial-dependent profile of external field. This phenomenon explicitly
demonstrates the exotic corpuscular and anisotropic property of polaritons.
\end{abstract}

\maketitle

\section{\label{sec:one}Introduction}

Quasi-particles are excitations of the matter. According to modern many body
theory, Elementary particles and quasi-particles are basic construction of
matters. The latter are crucial for understanding many phenomena in
condensed matter physics. Actually, quasi-particles can be regarded as
collective excitations of many elementary particles, as well as the mixtures
of different elementary excitations, whose behavior are similar to the matter
particles \cite{book1}.

In atomic physics and quantum optics, some exotic phenomena can be explained
by the concept of quasi-particles. For example, slow light phenomenon \cite%
{slow,slow397} in electromagnetically induced transparency (EIT) \cite%
{Harris,Harris82} can be explained in terms of quasi-particles - polaritons
\cite{Lukin84,Lukin65,Lukin77,Lukin75}. EIT happens when a weak signal light
field and a stronger control field are coupled to an ensemble of atoms with
a $\Lambda $ energy level configuration. Under the two-photon resonance, due
to the destructive interference between two interaction paths, the initially
opaque resonant medium becomes transparent with respect to the probe field,
and the group velocity of light is slowed down. Light then is stopped in the
EIT medium because only the dark state polariton is excited. The dark state
polariton is a bosonic like collective excitation, which is a mixture of a
signal light field and an atomic spin wave \cite{sprl91}.

Most recently, the light deflection was observed for the EIT atomic medium
in an external field with spatially inhomogeneous distribution \cite%
{Karpa,Scul07}. In the experiment of Ref. \cite{Karpa}, it is found that the
light ray bends when a magnetic field with small gradient vertical to the
propagation direction is applied to a cell with $\Lambda $-type rubidium
gas. This experiment was interpreted as the Stern-Gerlach experiment of the
dark polariton, thus the effective magnetic moment of the dark state
polariton is observed for the first time. It demonstrates that the dark
state polariton indeed behaves as a matter particle with mass, momentum and
magnetic moment et al, which can be reflected, refracted, and even deflected
by a gradient force. Therefore, quasi-particles show their particle nature
with definite momentum and effective mass. Different from that in Ref. \cite%
{Karpa}, the experiment of Re. \cite{Scul07} shows that a light can also
been deflected by an optical driven Rb atomic vapor when the profile of the
driving field is inhomogeneous. In this situation the angle of deviation is
an order of magnitude larger than that in Ref. \cite{Karpa}. The observed
phenomenon about light deflection in such EIT media has been explained
correctly according to the semi-classical theory \cite{ZDL} without using
the concept of dark-state polariton, which needs the quantization of light
fields.

Like matter particles, quasi-particles possess the wave-particle duality,
that is, quasi-particles sometimes appear to behave as particles and
sometimes appear to behave as waves. Here, we are interested in the particle
aspect of the dark polariton, which is an atomic collective excitation
dressed by the quantized probe light. The main purpose of
this paper is to systematically develop a quantum theory describing the
spatial motion of polaritons in inhomogeneous magnetic and optical fields.
We begin our investigation with the propagation of quasi-particles in the
limits of atomic linear response, where the atomic equations are treated
perturbatively. With an effective potential induced by the steady atomic
response in the external spatial-dependent field, the dynamics of spatial
motion of the quasi-particles is governed by the effective Schr\"{o}dinger
equation. The spatial motion of the quasi-particle is of anisotropic
depression -- the longitudinal motion is like a ultra-relativistic motion of
a ``slow light'' while the
transverse motion is of non-relativity with certain effective mass.

This paper is organized as follows: In sec. \ref{sec:two}, we present the
theoretical model for a $\Lambda $-type atomic ensemble in the presence of
inhomogeneous external fields, and derive the system of equations governing
the spatial motion of the signal field in the atomic linear response with
respect to the probe field. In Sec. \ref{sec:three}, the perturbation theory
is applied to obtain the atomic motion equation which is related to the
linear response to the signal field. In Sec. \ref{sec:four} and Sec. \ref%
{sec:five}, the crucial idea of the EIT - the dark-state polariton is
introduced as an dressed fields to describe the spatial motion of collective
excitation. Afterward, the dynamics of the quasi-particle - dark polariton
is discussed in the presence of an inhomogeneous magnetic field with a
spatial distribution along the transverse direction. In Sec. \ref{sec:six},
the spatial motion of the signal light in an inhomogeneous coupling field is
investigated. Then we make our conclusion in Sec. \ref{sec:sum}.

\section{\label{sec:two}theoretical model for $\Lambda$-type atomic ensemble
in external fields}

\begin{figure}[ptb]
\includegraphics[width=6 cm]{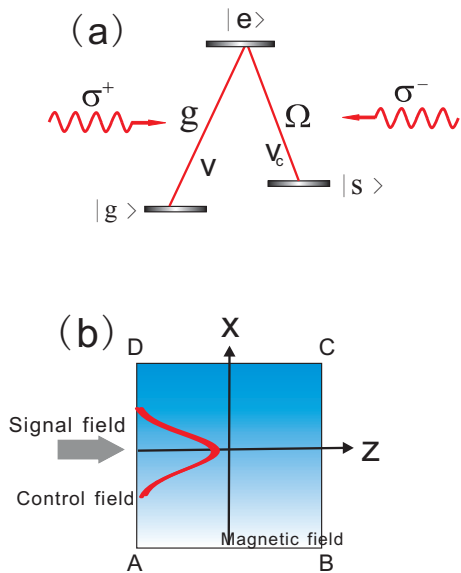}
\caption{\textit{(Color on line)} (a) Level scheme of atoms interacting with
$\protect\sigma^{+}$-polarized probe and $\protect\sigma^{-}$-polarized
probe strong fields. $\Omega$ denotes the Rabi frequency of the undepleted
classical control field. (b) Configuration of the optical beams and the
magnetic field inside the atomic medium.}
\label{fig:1}
\end{figure}

We consider an ensemble of $N$ identical and noninteracting atoms, which is
confined in a cell ABCD as shown in Fig.~\ref{fig:1}b. Each of the atoms is modeled
by a $\Lambda $-shaped energy level configuration with internal states $%
\left\vert g\right\rangle $, $\left\vert s\right\rangle $ and
$\left\vert e\right\rangle $. The transitions from the two lower
states $|g\rangle $ and $|s\rangle $ to the excited state $|e\rangle
$ are coupled by two optical fields, a weaker probe field and a
stronger control field, as shown in the top panel of Fig.
\ref{fig:1}a. The atomic transition from $\left\vert g\right\rangle
$ to $\left\vert s\right\rangle $ is forbidden by the electronic
dipole coupling. The probe field carries frequency $\nu $ and the
wave number $k$. It is a quantized electromagnetic field with
$\sigma ^{+}$ polarization. Under the rotating wave approximations,
its negative frequency part of the electric field
$\tilde{E}^{+}\left( \mathbf{r},t\right) $, couples the ground state
$|g\rangle $ to the excited state $|e\rangle $ at resonance, in the
absence of the magnetic field $B\left( \mathbf{r} \right) $. The
control field has carrier frequency $\nu _{c}=\omega
_{e}-\omega _{s}$ and wave number $k_{c}$. It is a classical field with $%
\sigma ^{-}$-polarized, and couples to the upper state $|e\rangle $ and the
metastable state $|s\rangle $ with Rabi frequency $\Omega \left( \mathbf{r}%
\right) $. After the magnetic field is applied along the
$z$-direction, the internal energies of the corresponding states are
shifted from their origins by magnitudes $\mu _{i}B$ with
\begin{equation}
\mu _{i}=m_{F}^{i}g_{F}^{i}\mu _{B},i\in \left\{ g,s,e\right\} .
\end{equation}%
Here, $\mu _{B}$ is the Bohr magneton, $g_{F}$ is the Land\'{e}
g-factor of the internal state $i$, and $m_{F}$ is the magnetic
quantum number.

As shown in Fig. \ref{fig:1}b, the probe field and the control field
propagate parallel in the $z-$ direction with wave number $k$ and
$k_{c}$ respectively. The Hamiltonian of this typical EIT system is
given by $H=H^{(A)}+H^{(F)}+H^{(I)}$. Let us use
$\tilde{\sigma}_{\mu \nu }^{j}\left( t\right) =\left\vert \mu
\right\rangle _{j}\left\langle \nu \right\vert $
to denote the internal state operator of the $j$-th atom between states $%
\left\vert \mu \right\rangle $ and $\left\vert \nu \right\rangle $. We
introduce the collective atomic operator \cite{Lukin65}
\begin{equation}
\tilde{\sigma}_{\mu \nu }\left( r,t\right) =\frac{1}{N_{r}}\sum_{r_{j}\in
N_{r}}\tilde{\sigma}_{\mu \nu }^{j}\left( t\right) \text{,}  \label{2-01a}
\end{equation}%
which is averaged over a small but macroscopic volume containing many atoms $%
N_{r}=\left( N/V\right) dV\gg 1$ around position $r$. Here $N$ is the number
of atoms in an interaction volume $V$. Then the Hamiltonian of the atomic
part reads%
\begin{equation}
H^{(A)}=\frac{N}{V}\sum_{j}\int d^{3}r\left( \omega _{j}-\mu _{j}B\right)
\tilde{\sigma}_{jj},  \label{2-01}
\end{equation}%
where we have neglected the kinetic term of atoms, and $\omega _{j}$
are the corresponding energy level spacing of the internal atomic
level. $H^{(F)} $ is the free Hamiltonian of the radiation field.
Using the electric-dipole approximation and the rotating-wave
approximation, the interaction with electromagnetic field reads
\cite{Lukin84,Lukin65,Lukin77,Lukin75}
\begin{equation}
H^{(I)}=-\frac{N}{V}\int d^{3}r\left[ d_{eg}\tilde{\sigma}_{eg}\tilde{E}%
^{+}+\Omega \tilde{\sigma}_{es}e^{i\left( k_{c}z-\nu _{c}t\right) }+h.c.%
\right]   \label{2-02}
\end{equation}%
Here, $\tilde{E}^{+}$ is the negative frequency of the probe field;
$\Omega \left( \mathbf{r}\right) $ is the Rabi frequency of the
control field, which usually depends on the spatial coordinate
through the spatial profile of
driving field; $d_{eg}$ is the dipole matrix element between the states $%
|g\rangle $ and $|e\rangle $.

For convenience, we describe the electric field as
\begin{equation}
\tilde{E}^{+}\left( r,t\right) =\sqrt{\frac{\nu }{2\varepsilon _{0}V}}%
E\left( r,t\right) e^{i\left( kz-\nu t\right) }  \label{2-03}
\end{equation}%
in the following discussion. Here $\exp \left[ i\left( kz-\nu
t\right) \right]$ is the carrier wave with frequency $\nu $ and wave
number $k$ propagating in the $z$ direction, and $E\left( r,t\right)
$ is the slow varying envelope, meaning that its spatiotemporal
variation is much slower than the carrier wave length and frequency.
Further we introduce the slowly varying variables for the atomic
transition operators
\begin{subequations}
\label{2-04}
\begin{align}
\tilde{\sigma}_{eg}\left( r,t\right) & =\sigma _{eg}\left( r,t\right)
e^{-ikz}\text{,} \\
\tilde{\sigma}_{es}\left( r,t\right) & =\sigma _{es}\left( r,t\right)
e^{-ik_{c}z}\text{.}
\end{align}%
In the rotating reference frame, the dynamics of this system is described by
the interaction Hamiltonian
\end{subequations}
\begin{align}
H_{I}& =-\frac{N}{V}\int d^{3}r[\sum_{j}\mu _{j}B\sigma _{jj}  \label{2-05}
\\
& +\left( g\sigma _{eg}E+\Omega \sigma _{es}+h.c.\right) ]\text{,}  \notag
\end{align}%
where the atom-field coupling constant $g$ is defined as
\begin{equation}
g=d_{eg}\sqrt{\frac{\nu }{2\varepsilon _{0}V}}.  \label{2-06}
\end{equation}

Before we study the EIT features of this system in detail, let us
first stand in the point of view of light to investigate the
propagation effects of pulses in an atomic medium. It is well-known
that, when atoms are subjected to an electric field, the applied
field displaces the positive charges and the negative charges in
atoms from their usual positions. This small movement that positive
charges in one direction and negative ones in the other will result
in collective induced electric-dipole moments. All dipole moments in
the dielectric material generate the polarization collectively,
which is defined as the collective dipole moment per unit volume
\begin{equation}
P=\frac{N}{V}d_{ge}\sigma _{ge}e^{i\left( kz-\omega _{eg}t\right) }+h.c.%
\text{.}  \label{2-07}
\end{equation}%
The collective dipole moment in Eq.(\ref{2-07}) is caused by the
atomic response to an optical electric field in a dielectric
material. In turn, every dipole with a nonvanishing second
derivative in time radiates an electromagnetic wave, that is, the
dielectric response $P$ of the medium acts as an effective source to
produce the electromagnetic field.

The Heisenberg equation for the slowly varying field operator $E\left(
r,t\right) $ results in a paraxial wave equation in classical optics \cite%
{Lukin84,Lukin65}
\begin{equation}
\left( i\frac{\partial }{\partial t}+ic\frac{\partial }{\partial z}+\frac{c}{%
2k}\nabla _{T}^{2}\right) E=-g^{\ast }N\sigma _{ge}\text{.}  \label{2-10}
\end{equation}%
Here, $c=1/\sqrt{\varepsilon _{0}\mu _{0}}$ is the velocity of light in
vacuum and the transverse Laplacian is defined as
\begin{equation}
\nabla _{T}^{2}=\partial ^{2}/\partial x^{2}+\partial ^{2}/\partial y^{2}
\end{equation}%
in the rectangular coordinates. When we neglect the $x$- and $y$-dependence
of $\tilde{E}$, that is, confine the problem in one dimension, Eq. (~\ref
{2-10}) immediately reduces to the usual propagation equation
\begin{equation}
\left( i\frac{\partial }{\partial t}+ic\frac{\partial }{\partial z}\right)
E=-g^{\ast }N\sigma _{ge}
\end{equation}%
given in Ref.~\cite{Lukin84,Lukin65,Lukin77,Lukin75,Sculb}, which only
describe light propagation in $z$ direction. To consider the problem in
three spatial dimensions, one can use the paraxial wave equation (~\ref{2-10}
)\ to investigate the dynamics of the input pulse in a resonant atomic
medium.

In this paper we will focus on the case that the linear optical
response theory works well, which can sufficiently reflect the main
physical features of the spatial motion of the input pulse with slow
group velocity. The lowest order contribution to the polarization is
the linear response of atoms defined as
\begin{equation}
P^{(1)}=\frac{N}{V}d_{ge}\sigma _{ge}^{(1)}e^{i\left( kz-\omega
_{eg}t\right) }\text{.}  \label{2-11}
\end{equation}%
Then, the paraxial wave equation becomes%
\begin{equation}
i\frac{\partial }{\partial t}E+ic\frac{\partial }{\partial z}E+\frac{c}{2k}%
\nabla _{T}^{2}E=-g^{\ast }N\sigma _{ge}^{(1)}\text{,}  \label{2-12}
\end{equation}%
where the definition of $\sigma _{ge}^{(1)}$ will be given in the next
section.

\section{\label{sec:three} perturbation approach}

We now study the evolution of the atomic ensemble under the
influence of the applied optical fields. The dynamics of this atomic
ensemble is described by the Heisenberg equations
\begin{subequations}
\label{3-01}
\begin{align}
\dot{\sigma}_{eg} & =\left[ i\left( \mu_{g}-\mu_{e}\right) B\right]
\sigma_{eg}-i\Omega^{\ast}\sigma_{sg} \\
& +ig^{\ast}\left( \sigma_{ee}-\sigma_{gg}\right) E^{+}\text{,}  \notag \\
\dot{\sigma}_{es} & =\left[ i\left( \mu_{s}-\mu_{e}\right) B\right]
\sigma_{es}-ig^{\ast}\sigma_{gs}E^{+} \\
& +i\Omega^{\ast}\left( \sigma_{ee}-\sigma_{ss}\right) \text{,}  \notag \\
\dot{\sigma}_{sg} & =\left[ i\left( \mu_{g}-\mu_{s}\right) B\right]
\sigma_{sg}-i\Omega\sigma_{eg}+ig^{\ast}\sigma_{se}E^{+}\text{.}
\end{align}
Since EIT is primarily concerned with the nonlinear modification of the
optical properties of the probe field, thus the low density approximation is
valid. In this approximation, the intensity of the quantum field is much
weaker than that of the coupling field $\Omega$, and the number of photons
contained in the signal pulse is much less than the number of atoms in the
sample.

In the low density approximation, the perturbation approach can be applied
to the atomic part, which is introduced in terms of perturbation expansion~\cite{Lukin84,Lukin65,Lukin77,Lukin75}
\end{subequations}
\begin{equation}
\sigma _{ij}=\sigma _{ij}^{(0)}+\lambda \sigma _{ij}^{(1)}+\lambda
^{2}\sigma _{ij}^{(2)}+\cdots \text{,}  \label{3-02}
\end{equation}%
where $i,j=\left\{ e,s,g\right\} $ and $\lambda $ is a continuously varying
parameter ranging from zero to unity. Here $\sigma _{ij}^{(0)}$ is of the
zeroth order in $gE$, $\sigma _{ij}^{(1)}$ is of the first order in $gE$ and
so on. We now substitute Eq. (\ref{3-02}) into Eq. (\ref{3-01}) and retain
only terms up to the first order in the signal field amplitude. We thereby
obtain the system of equations in the zeroth order
\begin{subequations}
\label{3-03}
\begin{align}
\dot{\sigma}_{eg}^{(0)}& =d_{1}^{\ast }\sigma _{eg}^{(0)}-i\Omega ^{\ast
}\sigma _{sg}^{(0)}\text{,} \\
\dot{\sigma}_{es}^{(0)}& =d_{3}^{\ast }\sigma _{es}^{(0)}+i\Omega ^{\ast
}\left( \sigma _{ee}^{(0)}-\sigma _{ss}^{(0)}\right) \text{,} \\
\dot{\sigma}_{sg}^{(0)}& =d_{2}^{\ast }\sigma _{sg}^{(0)}-i\Omega \sigma
_{eg}^{(0)}\text{.}
\end{align}%
where the parameters
\end{subequations}
\begin{subequations}
\label{3-11}
\begin{align}
d_{1}& =i\left( \mu _{e}-\mu _{g}\right) B-\gamma _{1}\text{,} \\
d_{2}& =i\left( \mu _{s}-\mu _{g}\right) B-\gamma _{2}\text{,} \\
d_{3}& =i\left( \mu _{e}-\mu _{s}\right) B-\gamma _{3}\text{.}
\end{align}
and we have phenomenologically introduced the energy-level decay rates $%
\gamma _{i}$( $i\in \left\{ 1,2,3\right\} $).

We assume that all population of atoms are initially prepared in the ground
state $\left\vert g\right\rangle $ in the absence of electromagnetic fields,
and the depletion of the ground state is not significant for any time $t>0$
due to the quantum interference effect, therefore
\end{subequations}
\begin{equation}
\sigma _{gg}^{(0)}=1  \label{3-08}
\end{equation}%
while others vanish~\cite{Lukin65}. Then, the first order atomic
transition operator $\sigma _{ij}^{(1)}$, which are related to the
atomic linear response to the probe field, satisfy the following
equations~\cite{Lukin65}
\begin{subequations}
\label{3-09}
\begin{align}
\dot{\sigma}_{eg}^{(1)}& =d_{1}^{\ast }\sigma _{eg}^{(1)}-i\Omega ^{\ast
}\sigma _{sg}^{(1)}-ig^{\ast }E^{\dag }\text{,} \\
\dot{\sigma}_{es}^{(1)}& =d_{3}^{\ast }\sigma _{es}^{(1)}+i\Omega ^{\ast
}\left( \sigma _{ee}^{(1)}-\sigma _{ss}^{(1)}\right) \text{,} \\
\dot{\sigma}_{sg}^{(1)}& =d_{2}^{\ast }\sigma _{sg}^{(1)}-i\Omega
\sigma _{eg}^{(1)}\text{.}
\end{align}%
In order to get the equations of motion for polaritons, we rewrite
Eq.~(\ref{3-09}) as~\cite{Lukin65}
\end{subequations}
\begin{subequations}
\label{3-10}
\begin{align}
gE& =-\left[ \left( \partial _{t}-d_{1}\right) \frac{1}{\Omega }\left(
\partial _{t}-d_{2}\right) +\Omega \right] \sigma _{gs}^{(1)}\text{,} \\
\sigma _{ge}^{(1)}& =-\frac{i}{\Omega }\left( \partial _{t}-d_{2}\right)
\sigma _{gs}^{(1)}\text{,}
\end{align}

Equations (\ref{2-12}) and (\ref{3-10}) constitute a self-consistent system
of equations, which indicates that the polarization field $\sigma_{ge}^{(1)}
$ can serve as a source to generate the electric fields, whereas the
propagating light in turn drives the atomic media via the dipole
interaction. They are the starting point of our investigation in the
following several sections, where we study the phenomena of light deflection
that occur as a consequence of the interaction between the $\Lambda$-type
atomic ensemble and an external field with a spatial distribution.

\section{\label{sec:four}Spatial motion of quasi-particle in a harmonic magnetic field}

It is well known that the EIT system has two remarkable properties:
1) the opaque absorption medium becomes transparent with respect to
the probe light at certain frequencies. It happens because the
absorption on both transitions is suppressed by the destructive
interference between the excitation pathways to the upper level.
Thus a transparency window is rendered over a narrow spectral range
within the absorption line. 2) The group velocity of the incoming
pulse has been largely reduced within the transparency window.
Physically, the slow light in a EIT system is interpreted by the
formation of so called dark-state polariton (DSP). A ``dark-state
polariton'' is a bosonic-like collective excitation of a signal
light field and an atomic spin wave
\cite{Lukin84,Lukin65,Lukin77,Lukin75}, whose relative amplitude is
determined by the control laser field. In this section, we study the
dynamic of the DSP in the presence of a harmonic field with a
spatially inhomogeneous distribution in the transverse direction,
where the control field is assumed to be independent of position and
time.

When the light pulse enters a medium, photons interact with atoms of
the medium. They then combine together to form a type of excitations
known as polaritons, which are one kind of quasi-particles. In an
EIT system, two types of polaritons are introduced - the dark
polariton and the bright polariton, which are described respectively
by the dark polariton field operator $\Psi$ and bright polariton
field operator $\Phi$
\end{subequations}
\begin{subequations}
\label{4-01}
\begin{align}
\Psi\left( r,t\right) & = E\cos\theta-\sqrt{N}\sigma_{gs}^{(1)}\sin \theta%
\text{,} \\
\Phi\left( r,t\right) & = E\sin\theta+\sqrt{N}\sigma_{gs}^{(1)}\cos \theta%
\text{.}
\end{align}
They are atomic collective excitation (quasi-spin wave) dressed by the
quantized probe light with the inverse relations
\end{subequations}
\begin{subequations}
\label{4-02}
\begin{align}
E & = \Psi\cos\theta+\Phi\sin\theta \\
\sigma_{gs}^{(1)} & = \frac{1}{\sqrt{N}}\left( \Phi\cos\theta-\Psi\sin
\theta\right) \text{.}
\end{align}
The dark polariton field operator $\Psi$ and bright polariton field operator
$\Phi$ have bosonic commutation relations in the limit of few photons and
many atoms. And the action of $\Psi^{\dag}$ on the vacuum creates the dark
states, which contain no component of the excited state $\left\vert
e\right\rangle $. By assuming that the Rabi frequency is real, the mixing
angle of the signal field and the collective atomic polarization is given by
\end{subequations}
\begin{equation}
\tan\theta=\frac{g\sqrt{N}}{\Omega}\text{,}  \label{4-03}
\end{equation}
where the Rabi frequency $\Omega$ is related to the control laser
power $P$ \ through $\Omega^{2}=2\left\vert d_{es}\right\vert
^{2}P/\left( c\epsilon_{0}S\right) $. In the nearly two-photon
resonant condition, the only excitations are dark polaritons, which
generate an eigen-state with vanishing eigenvalue of the interaction
Hamiltonian. It is found from Eq.~(\ref{4-01}) and (~\ref{4-03})
that, by reducing the amplitude of the control field, the
contributions of light or atoms to the DSP can be changed, then the
DSP varies from photons to atoms. Thus it is roughly seen that the
mixing angle $\theta$ determines whether or not the group velocity
of the signal pulse propagating in the atomic medium can be
decreased.

In order to find how the mixing angle affects the group velocity of
the input pulse, we derive the equations of the spatial motion for
the dark polariton fields. In the above sections, we have achieved
the dynamic motion equations of atoms and light. In terms of the
field operators for the dark and bright polaritons, Eq.~(\ref{3-10})
and (~\ref{2-12}) can be rewritten as
\begin{align}
& g\sqrt{N}\left( \Psi \cos \theta +\Phi \sin \theta \right) =  \label{4-04}
\\
& -[\left( \partial _{t}-d_{1}\right) \frac{1}{\Omega }\left( \partial
_{t}-d_{2}\right) +\Omega ]  \notag \\
& \times \left( \Phi \cos \theta -\Psi \sin \theta \right)   \notag
\end{align}%
and
\begin{align}
& \left( i\frac{\partial }{\partial t}+ic\frac{\partial }{\partial z}+\frac{c%
}{2k}\nabla _{T}^{2}\right) \left( \Psi \cos \theta +\Phi \sin \theta
\right) =  \notag \\
& -\frac{g\sqrt{N}}{i\Omega }\left( \partial _{t}-d_{2}\right) \left( \Phi
\cos \theta -\Psi \sin \theta \right) .  \label{4-05}
\end{align}%
For a very small magnetic fields, we have $\left\vert \mu _{g}-\mu
_{e}\right\vert \ll \gamma _{1}$. Furthermore, we assume a
sufficiently strong driving field such that $\Omega ^{2}\gg \gamma
_{1}\gamma _{2}$. In the adiabatic approximation, the excitation of
the bright polariton field $\Phi$ vanishes approximately. Then the
dynamics of the dark polariton field $\Psi$
is governed by the Schr\"{o}dinger-like equation%
\begin{equation}
i\frac{\partial }{\partial t}\Psi =[\check{T}+V\left( r\right) ]\Psi
\label{4-06}
\end{equation}%
with an effective potential
\begin{equation}
V(r)=-\left( \mu _{s}-\mu _{g}\right) B\left( r\right) \sin ^{2}\theta
\end{equation}%
induced by the steady atomic response in the external spatial-dependent
field. Here, we have set $\gamma _{2}=0$; while the effective kinetic
operator%
\begin{equation}
\check{T}=v_{g}p_{z}-\frac{1}{2k}\cos ^{2}\theta \nabla _{T}^{2}
\label{4-06-b}
\end{equation}%
represents an anisotropic depression, where the momentum along z
direction is defined as $p_{z}\equiv -i\partial _{z}$. The
longitudinal term $v_{g}p_{z}$ in Eq.(\ref{4-06-b}) describes a
ultra-relativistic motion with a slow light velocity
\begin{equation}
v_{g}=c\cos ^{2}\theta \text{,}  \label{4-07}
\end{equation}%
while the transverse part $P_{T}^{2}/(2m)$ in the effective kinetic term
desribes a non-relativistic motion with an effective transverse mass

\begin{equation}
m=\frac{k}{v_{g}}=\frac{k}{c}\sec^{2}\theta.
\end{equation}
The above effective Schr\"{o}dinger equation governs the dynamics of spatial
motion of quasi-particles.

Obviously, when no magnetic field is applied, due to the transverse
Laplacian operator $\nabla _{T}^{2}$ commutating with $\partial _{z}$, we
can separate the $z$-component from $x$-$y$ component. Neglecting the $x$-
and $y$-dependence of $\Psi $, Eq. (\ref{4-06}) describes a stable
propagation along $z$- axis with group velocity $v_{g}$ \cite%
{Harris,Harris82,Lukin84,Lukin65,Lukin77,Lukin75}. Hence, the amplitude of
the control field determines the group velocity of the input pulse in the
atomic medium. Adiabatically rotating the angle from $0$ to $\pi /2$, the
polariton can be decelerated to a full stop. On the other hand by increasing
the strength of the coupling field, that is, reversing the rotation of $%
\theta $ adiabatically, it leads to a re-acceleration of the dark-state
polariton associated with a change of character from collective spin-like
waves to electromagnetic photons.

Now we consider the three-dimension problem. By defining $P_{j}\equiv
-i\partial _{j}$( $j\in \left\{ x,y,z\right\} )$, Eq.(\ref{4-06}) can be
rewritten as an effective Schr\"{o}dinger equation%
\begin{equation}
i\frac{\partial }{\partial t}\Psi =H_{eff}\Psi   \label{4-08}
\end{equation}%
with the effective Hamiltonian
\begin{equation}
H_{eff}=v_{g}P_{z}+\frac{1}{2m}\left( P_{x}^{2}+P_{y}^{2}\right) -\mu
^{\prime }B\left( r\right) \text{,}  \label{4-09}
\end{equation}%
where $\mu ^{\prime }=\left( \mu _{s}-\mu _{g}\right) \sin ^{2}\theta $. The
magnitude of the effective transverse mass is totally determined by the
mixing angle $\theta $ of the signal field and the collective atomic
polarization. When the amplitude of the control field is small, spin waves
have large contributions to the DSP, therefore the effective transverse mass
is large; when the Rabi frequency is large, photons give large contributions
to the DSP, therefore the effective transverse mass is small. The effective
Schr\"{o}dinger equation (\ref{4-08}) is the starting point for
investigating the spatial motion of the dark polariton. It shows that, due
to the inhomogeneity of the magnetic field, the motion of the dark polariton
will be scattered by an effective potential with value $\mu ^{\prime
}B\left( r\right) $.

Now we assume that the magnetic field in $z$ direction has a spatial
distribution in the transverse direction with the expression
\begin{equation}
B\left( r\right) =B_{0}+B_{x}x^{2}+B_{y}y^{2}\text{,}  \label{4-10}
\end{equation}
where $B_{x},B_{y}<0$. Then the effective Hamiltonian operator
becomes $ H_{1}=H_{ma}+v_{g}P_{z}-\mu^{\prime}B_{0}$ with
\begin{equation}
H_{ma}=\frac{P_{x}^{2}}{2m}+\frac{m\omega_{x}^{2}}{2}x^{2}+\frac{P_{y}^{2}}{%
2m}+\frac{m\omega_{y}^{2}}{2}y^{2}.  \label{4-11}
\end{equation}
In classical physics, $H_{ma}$ corresponds to a two dimensional
harmonic oscillator with mass $m$ and angular frequency
$\omega_{x}=\sqrt{-2\mu ^{\prime}B_{x}/m}$ in $x$-direction and
$\omega_{y}=\sqrt{ -2\mu^{\prime}B_{y}/m}$ in $y$-direction. For a
given initial state $ \Psi\left( 0\right) $, the evolution state
$\Psi\left( t\right)$ of the system is a unitary transformation of
the initial state $\Psi\left( 0\right) $ with the time-evolution
unitary operator $U\left( t\right) =U_{z}\left( t\right) U_{y}\left(
t\right) U_{x}\left( t\right) $:
\begin{subequations}
\label{4-12}
\begin{align}
U_{z} & =\exp\left( -iv_{g}P_{z}t\right) \text{,} \\
U_{y} & =\exp\left[ -i\left( \frac{P_{y}^{2}}{2m}+\frac{m\omega_{y}^{2}}{2}%
y^{2}\right) t\right] \text{,} \\
U_{x} & =\exp\left[ -i\left( \frac{P_{x}^{2}}{2m}+\frac{m\omega_{x}^{2}}{2}%
x^{2}\right) t\right] \text{.}
\end{align}

Next we consider the evolution dynamics of a spatially well-localized wave
packet, which is centered at $(x_{0},y_{0},z_{0})=(0,0,0)$ and has a
vanishing mean velocity in all directions. The spatially well-localized wave
packet is assumed to be initially in a Gaussian form
\end{subequations}
\begin{equation}
\Psi \left( 0\right) =\prod\limits_{\xi =x,y,z}\left( \frac{\alpha _{\xi }}{%
\pi }\right) ^{1/4}e^{-\frac{1}{2}\alpha _{\xi }\xi ^{2}}  \label{4-13}
\end{equation}%
with width $1/\sqrt{\alpha _{\xi }}$, $\xi \in \left\{ x,y,z\right\} $ in $%
x,y,z$-direction, respectively, and $\alpha _{\zeta }\neq \lambda
_{j}$ where $\lambda _{j}=m\omega _{j}$. By getting rid of an
irrelevant global phase factor, the wave function at time $t>0$
reads
\begin{align}
\Psi \left( t\right) & =\left[ \frac{\alpha _{z}}{\pi }\right] ^{1/4}e^{-%
\frac{1}{2}\alpha _{z}\left( z-v_{g}t\right) ^{2}}  \label{4-14} \\
& \sum_{n_{1}n_{2}}C_{2n_{1}}^{(x)}C_{2n_{2}}^{(y)}e^{-i2\left( n_{1}\omega
_{x}+n_{2}\omega _{y}\right) t}\phi _{2n_{1}}^{(x)}\phi _{2n_{2}}^{(y)}\text{%
.}  \notag
\end{align}%
The initial Gaussian wave packet evolves into a superposition of the product
states $\phi _{n_{1}}^{(x)}\phi _{n_{2}}^{(y)}$ with quantum numbers $n_{1}$
and $n_{2}$ taking on the value $0,1,2\cdots $. Coefficients in Eq.(\ref%
{4-14}) read
\begin{equation}
C_{2n}^{(j)}=\frac{\sqrt{\left( 2n\right) !}}{2^{n}n!}\left[ \frac{4\lambda
_{j}\alpha _{j}}{\left( \lambda _{j}+\alpha _{j}\right) ^{2}}\right]
^{1/4}\left( \frac{\lambda _{j}-\alpha _{j}}{\lambda _{j}+\alpha _{j}}%
\right) ^{n}\text{.}  \label{4-15}
\end{equation}%
Here, $\phi _{n}^{(x)}$ is the eigenfunction of Hamiltonian $%
P_{x}^{2}/\left( 2m\right) +m\omega _{x}^{2}x^{2}/2$ with the corresponding
eigenvalue $E_{n}=\left( n+1/2\right) \omega _{x}$
\begin{equation}
\phi _{n}^{(x)}=\left[ \frac{1}{2^{n}n!}\right] ^{\frac{1}{2}}\left( \frac{%
\lambda _{x}}{\pi }\right) ^{\frac{1}{4}}H_{n}\left( \sqrt{\lambda _{x}}%
x\right) e^{-\frac{1}{2}\lambda _{x}x^{2}}\text{,}  \label{4-16}
\end{equation}%
where $H_{n}\left( x\right) $ are Hermite polynomials. The wave function $%
\phi _{n}^{(y)}$ has the similar expression as Eq.(\ref{4-16}) with $x$
replaced by $y$.

\begin{figure}[ptb]
\includegraphics[width=6 cm]{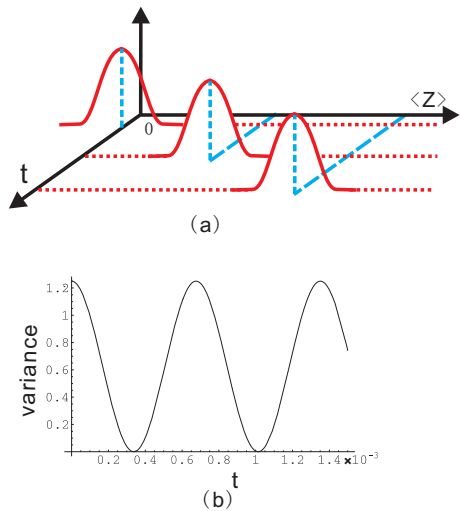}
\caption{\textit{(Color on line)} (a) Schematic illustration about the time
evolution for the center of the initial Gaussian state of the dark polariton
along $z$ direction. (b) the time evolution of the variance in $x$%
-direction. The variance is in unit of $0.1$. }
\label{fig:2}
\end{figure}

Since the wave-function at time $t$ is an even function of variables
$x$ and $y$, which can be seen from Eq. (\ref{4-14}) to Eq.
(\ref{4-16}), the trajectory of the center of the wave packet in the
$x$ - $y$ plane does not change with time. But in $z$-direction, the
dark polariton propagates with mean velocity $v_{g}$, and the center
of the wave packet leaves its original place proportionally to the
time $t$ with $\left\langle z\right\rangle =v_{g}t$. Although the
wave packet keeps it shape along $z$ direction with an unchanged
variance
\begin{equation}
\left( \Delta z\right) ^{2}=\frac{1}{4\alpha_{z}}\text{,}  \label{4-18}
\end{equation}
the variances in the $x$- and $y$-direction oscillate with time, namely,
\begin{align}
\left( \Delta x\right) ^{2} & = A_{-x}\cos\left( 2\omega_{x}t\right) +A_{+x}%
\text{,}  \label{4-17} \\
\left( \Delta y\right) ^{2} & = A_{-y}\cos\left( 2\omega_{y}t\right) +A_{+y}%
\text{,}  \notag
\end{align}
where
\begin{equation}
A_{\pm s}=\frac{2\pi
m^{2}\omega_{s}^{2}\pm\alpha_{s}^{2}}{8\alpha_{s}\pi
m^{2}\omega_{s}^{2}},s=x,y.
\end{equation}
Fig. \ref{fig:2} (a) schematically illustrates the time evolution of the
initial Gaussian packet. The wave packet distributed along the $z$-direction
keeps its original shape. However, in $x$-direction, the shape of the
Gaussian packet oscillates as time evolution, its width change is shown in
Fig. \ref{fig:2} (b).

In plotting Fig. \ref{fig:2} (b), we have taken the reasonable parameters
accessible in current expriments\cite{Karpa}: the width of the initial
Gaussian $1/\alpha_{x}=1/4$cm, the atomic density $N/V=10^{13}$ per cube
centimeter, the quantum number of the ground state $m_{F}^{g}=-2$ with $%
g_{F}^{g}=1/2$, the quantum number of the mestable state $m_{F}^{s}=0$, and $%
\gamma_{1}=2\pi\times2.87$MHz.

\section{\label{sec:five}The deflection of dark polaritons in a linear
magnetic field}

In a recent experiment \cite{Karpa}, a magnetic field with small
transverse gradient is applied to a $\Lambda$-type atomic medium. It
was observed that the light beam is deflected after the signal light
passing through the EIT gas cell. This observation first
demonstrates that quasi-particles - dark-state polaritons have a
non-zero magnetic moment. This experimental observation can be
interpreted straightforwardly according to the above quantum theory
of spatial motion of polaritons in inhomogeneous fields.

For simplicity, we assume the magnetic field
\begin{equation}
B\left( r\right) =B_{0}+B_{1}x\text{.}  \label{5-01}
\end{equation}
has a linear gradient along the $x$-direction. We further allow the input
pulse to vary only in one transverse dimension, says, in the $x$-direction,
which means that we neglect the $y$-dependence of the input pulse $E$. Then
the two dimensional effective Hamiltonian for the dark polaritons reads
\begin{equation}
H_{2}=v_{g}P_{z}+\frac{1}{2m}P_{x}^{2}-\mu b_{0}-\mu\zeta x\text{,}
\label{5-02}
\end{equation}
where the parameters
\begin{subequations}
\label{5-03}
\begin{align}
\zeta & = 2B_{1}\sin^{2}\theta\text{,} \\
b_{0} & = 2B_{0}\sin^{2}\theta\text{,} \\
\mu & = \mu_{s}-\mu_{g}
\end{align}
can be controlled by the mixing angle $\theta$. For an initial dark
polariton field with Gaussian spatial distribution
$\alpha_{x}=\alpha_{z}=b^{-2}$ as given in Eq.~(\ref{4-13}), the
time evolution of the polariton field reads $U_{li}\left( t\right)
=\exp\left( -iH_{2}t\right) $. By the Wei-Norman algebraic method
(see the appendix A) \cite{swna}, the unitary operator $U_{li}\left(
t\right) $ can be factorized as $U_{li}\left( t\right) =U_{2}\left(
t\right) U_{1}\left( t\right) $ with
\end{subequations}
\begin{subequations}
\label{5-04}
\begin{align}
U_{2}\left( t\right) & = e^{-iv_{g}tP_{z}}e^{-i\frac{t}{2m}%
P_{x}^{2}}e^{it^{2}\frac{\mu\zeta}{2m}P_{x}}\text{,} \\
U_{1}\left( t\right) & = e^{it\mu\zeta x}e^{i\mu b_{0}t-i\frac{t^{3}}{3}%
\frac{\mu^{2}\zeta^{2}}{2m}}\text{.}
\end{align}

A straight forward calculation shows that, the initial Gaussian packet
evolves into
\end{subequations}
\begin{align}
\Psi \left( t\right) & =\left( \frac{1/\pi }{b^{2}+i\frac{t}{m}}\right) ^{%
\frac{1}{2}}e^{i\mu t\left( b_{0}-\frac{t^{2}}{3}\frac{\mu \zeta ^{2}}{2m}%
\right) }e^{-\frac{\left( z-v_{g}t\right) ^{2}}{2b^{2}}}  \label{5-05} \\
& e^{i\mu \zeta tx}\exp \left[ -\frac{\left( x-t^{2}\frac{\mu \zeta }{2m}%
\right) ^{2}\left( b^{2}-i\frac{t}{m}\right) }{2b^{4}+2t^{2}/m^{2}}\right]
\text{.}  \notag
\end{align}%
At time $t$, the center of the input pulse moves to $v_{g}t$ along the $z$%
-direction and $t^{2}\mu \zeta /\left( 2m\right) $ along the
$x$-direction. When a dark polariton is excitated by the interaction
between light and atoms, the dark polariton will achieve a velocity
along the $x$-direction as
\begin{equation}
v_{x}=\frac{\mu \zeta L}{mv_{g}}\text{.}  \label{5-06}
\end{equation}%
after it pass through the gas cell with length L. Therefore the deflection
angle reads
\begin{equation}
\alpha =\frac{v_{x}}{v_{g}}=\frac{L}{v_{g}}\frac{\mu }{k}B_{1}\sin
^{2}\theta \text{.}  \label{5-07}
\end{equation}%
In real experiment, the dephasing time is nonzero due to the
collision between atoms, which leads to the absorption of the energy
of the probe beam by the atomic medium.

The above results mean that the deflection angle of the output pulse depends
on the mixing angle $\theta$ between the signal field and the collective
atomic polarization, the wave number $k$ of the input pulse, and the
gradient $B_{1} $ of the inhomogeneous magnetic field. One can find that the
magnetic moment of the dark polariton has an effective value
\begin{equation}
\mu_{pol}=\mu\sin^{2}\theta\text{.}  \label{5-08}
\end{equation}
By taking $m_{g}=-2$ and $m_{s}=0$, we find the effective magnetic moment
\begin{equation}
\mu_{pol}=2g_{F}^{(g)}\mu_{B}\sin^{2}\theta\text{,}  \label{5-08-0}
\end{equation}
which is exactly the theoretical result given in Ref. \cite{Karpa}.

Next we consider the spatial resolution, which in optics reflects the
ability of this optical system to form separate and distinct images of two
objects.
\begin{figure}[tbp]
\includegraphics[width=6 cm]{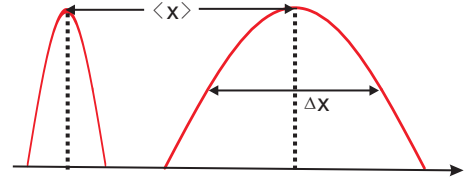}
\caption{\textit{(Color on line)}Schematic illustration about the spatial
resolution.}
\label{fig:3ad}
\end{figure}
The spatial resolution is defined here as the mean signal divided by its
standard deviation
\begin{equation}
R=\frac{\left\langle x\right\rangle }{\Delta x}=t^{2}\mu \zeta \sqrt{\frac{%
b^{2}}{2m^{2}b^{4}+2t^{2}}}\text{,}  \label{5-08-1}
\end{equation}%
where the mean position $\left\langle x\right\rangle $ in the transverse
direction and the standard deviation $\Delta x$ are given by
\begin{subequations}
\label{5-08-2}
\begin{align}
\left\langle x\right\rangle & =\int_{-\infty }^{+\infty }\Psi ^{\ast }\left(
t\right) x\Psi \left( t\right) dxdz=t^{2}\frac{\mu \zeta }{2m}\text{,} \\
\Delta x& =\sqrt{\left\langle x^{2}\right\rangle -\left\langle
x\right\rangle ^{2}}=\sqrt{\frac{m^{2}b^{4}+t^{2}}{2b^{2}m^{2}}}\text{.}
\end{align}%
It can be seen that the spatial resolution increases as the
interaction time between light and atoms increases.

Actually, the phenomenon of light deflection in such an
inhomogeneous magnetic field can also be described without using the
concept of quasiparticles - dark polaritons. Here, we show how to
calculate the deflection angle $\alpha $ in Eq. (\ref{5-07})
according to the semiclassical theory. We begin the semiclassical
approaches by considering the evolution of the system from Eq.
(\ref{2-12}) and (\ref{3-09}). First, the atomic linear response to
the signal field has been explicitly reflected in Eq. (\ref{3-09}).
Under the adiabatic approximation that the evolution of the atomic
system is much faster than the temporal change of the radiation
field, we can obtain the steady-state solution for the atomic
transition $\sigma _{ge}^{(1)}$ by setting all time derivatives to
zero in Eq. (\ref{3-09}), namely
\end{subequations}
\begin{align}
\sigma _{ge}^{(1)}& =i\frac{\left[ i\left( \mu _{g}-\mu _{s}\right) B+\gamma
_{2}\right] g}{d_{1}d_{2}+\left\vert \Omega \right\vert ^{2}}E  \label{5-09}
\\
& \approx \frac{g}{\left\vert \Omega \right\vert ^{2}}\mu BE.  \notag
\end{align}%
Here, the undepleted control-field approximation is used and $\gamma
_{2}\approx 0$ is assumed. This approach based on the atomic linear
response results in an effective potential for the motion of signal
slow-varying amplitude due to the spatial distribution of the
magnetic field. The spatial motion of the slow varying amplitude is
described by the following equation:
\begin{align}
& i\frac{\partial }{\partial t}E+ic\frac{\partial }{\partial z}E+\frac{c}{2k}%
\nabla _{T}^{2}E=-\frac{|g|^{2}N}{\left\vert \Omega \right\vert ^{2}}\mu BE
\label{5-10} \\
& =-\mu \left( B_{0}+B_{1}x\right) E\tan ^{2}\theta   \notag
\end{align}
which describes a shape-preserving propagation in $z$ direction with
velocity $c$.

For an initial Gaussian wave packet of $E$ in $x-z$ plane, after
passing through the gas cell, the wave center shifts from $\left(
x,z\right) =\left( 0,0\right) $ to the well-defined position
\begin{equation}
\left( x,z\right) =\left( \frac{\mu B_{1}}{2kc}L^{2}\tan^{2}\theta ,L\right)
\text{.}  \label{5-11}
\end{equation}
And it can be obviously found that Eq. (\ref{5-07}) is the deflection angle
in this approach.

\section{\label{sec:six}Deflection of light in inhomogeneous coupling field}

In this section, we turn our discussion to the deflection of slow
light in the atomic medium driven by a optical field with an
inhomogeneous profile, while the magnetic field is uniform. This
phenomenon was experimentally observed in Ref. \cite{Scul07} where
the cell filled with EIT-based atomic gas is referred as an
ultra-dispersive optical prism with an angular dispersion. We note
that the probe light is relatively strong in comparison with the
control light in this experiment, thus the susceptibility obtained
from the linear response theory can not work well to explain the
experiment phenomenon. In this paper, we do not plan to explain the
experiment data in Ref. \cite{Scul07} in the strong coupling limit.
Our main purpose is to predict a new quantum coherent phenomenon for
the light deflection by the atomic media when the experiment is
carried out for the weak probe field.

We assume the strong driving field has a Gaussian profile
\begin{equation}
\Omega=\Omega_{0}\exp\left[ -\frac{x^{2}}{2\sigma^{2}}\right]  \label{6-01}
\end{equation}
in the transverse direction. Here, we confine the problem to two dimensional
space, the $x$-$z$ plane. Then the transverse Laplacian operator reduces
into a one dimensional operator $\nabla_{T}^{2}=\partial^{2}/\partial x^{2}$.

By invoking the steady-state conditions, it is found that the
polarization field $\sigma_{ge}^{(1)},$ which serves as a source for
the electric fields in Eq. (\ref{2-12}), is proportional to the slow
varying amplitude $E$ given in Eq. (\ref{5-09}). Under a strong,
undepleted driving field approximation, the coupling between atoms
and light induces a spatial dependent potential into the propagation
equation. The spatial shape of this potential induced by
$\sigma_{ge}^{(1)}$ is completely determined by the profile of the
Rabi frequency $\Omega$, which can be seen in the first line of Eq.
(\ref{5-10} ). Thus, when the signal pulse parallel to the control
beam travels across the atomic cell, it will be scattered by the
effective potential. However, as the width of probe beam is less
than that of the control beam, the trajectory of the probe light may
bend when it is adjusted to the left side or to the right side of
the control beam profile; hence the probe and control beams are no
longer parallel after they go through the gas cell.

In order to investigate this phenomenon, we assume the probe beam is in a
Gaussian state
\begin{equation}
E\left( 0,x,z\right) =\frac{1}{\sqrt{\pi b^{2}}}\exp\left[ -\frac{\left(
x-a\right) ^{2}}{2b^{2}}-\frac{z^{2}}{2b^{2}}\right]  \label{6-02}
\end{equation}
before it enters the gas cell, where $b$($<\sigma$) is the width of the
probe field and $a$ is the initial location of the wave packet center of the
probe field along $x$-direction. The sign of $a$ indicates the incident
position comparatively to the left or right hand side of the control beam's
center $x_{0}=0$, and the magnitude $\left\vert a\right\vert $ denotes the
distance from the control beam's center. In order to investigate the
evolution of this initial state, we expand $\left\vert \Omega\right\vert
^{-2}$ at the position $a$ and retain the linear term proportional to $x-a$.

With the above considerations, the paraxial equation in Eq. (\ref{5-10})
becomes
\begin{equation}
i\dot{E}+ic\partial _{z}E+\frac{c}{2k}\partial _{x}^{2}E=\left( \eta
_{0}+\eta _{1}x\right) E  \label{6-03}
\end{equation}%
where
\begin{subequations}
\label{6-04}
\begin{align}
\eta _{0}& =-\Omega _{0}^{-2}\left( 1-\frac{2a^{2}}{\sigma ^{2}}\right)
|g|^{2}N\Delta \exp \left( \frac{a^{2}}{\sigma ^{2}}\right) \text{,} \\
\eta _{1}& =-2a\Delta \frac{|g|^{2}N}{\sigma ^{2}}\Omega _{0}^{-2}\exp
\left( \frac{a^{2}}{\sigma ^{2}}\right) \text{.}
\end{align}%
and $\Delta =\left( \mu _{s}-\mu _{g}\right) B$. By making use of
the Wei-Norman algebraic method \cite{swna}, it is shown that, after
passing through the Rb gas cell, the center position $\left(
x,z\right) =\left( a,0\right) $ of the probe field at time $t=0$ is
shifted to
\end{subequations}
\begin{subequations}
\label{6-05}
\begin{align}
x& =a+L^{2}\Omega _{0}^{-2}\Delta ae^{\frac{a^{2}}{\sigma ^{2}}}\frac{%
|g|^{2}N}{\sigma ^{2}kc}\text{,} \\
z& =L\text{.}
\end{align}

If we track the motion of the center of the probe beam, a mirage
effect occurs. The sign of $\Delta $ and the incident position $a$
of the signal light determine whether the trajectory of probe beam
bend. When the magnetic field is absent $\Delta =0$ or the center of
the probe field is collinear to that of the control field
$a=x_{0}=0$, the trajectory of the signal light is a straight line.
We assign the positive sign for $a$ as the probe beam is shifted to
the right with respect to the center of the control light, and
denote $a<0$ as the signal beam is shifted to the left. When the
probe beam is shifted to the right, in the case of $\Delta <0$, the
signal light feels a ``repulsion potential'' due to the coefficient
$\eta _{1}>0$, thus the trajectory bends to the left; at the
condition $\Delta >0$, the signal light undergoes an ``attractive
potential'' in the atomic medium due to $\eta _{1}<0$, thus the
trajectory bends to the right. When $a<0$, it can be found from Eq.
(\ref{6-04}) that due to the coefficient of the linear potential
larger than zero, i.e. $\Delta >0$, the probe beam experiences a
``repulsion potential'' within the EIT medium, and its center is
shifted to the left. As $\eta _{1}$ is smaller than zero, i.e.
$\Delta <0$, the probe beam suffers an ``attractive potential''
during its passing through the EIT medium, hence its center is
shifted to the right. The corresponding schematic diagram is given
in Fig.\ref{fig:3}, where yellow solid line is the spatial
distribution of the
\begin{figure}[tbp]
\includegraphics[width=6 cm]{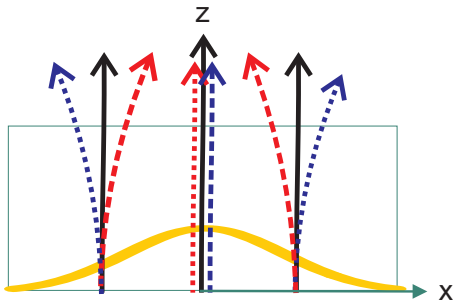}
\caption{\textit{(Color on line)}Schematic illustration about the ray
deflection of the probe light in the presence of inhomogenous coupling
light. The yellow line is the spatial distribution of the control light.}
\label{fig:3}
\end{figure}
control light, the red dash lines give the deflection at $\Delta <0$, the
blue dotted lines describe the light trajectory at $\Delta >0$ and the black
solid lines depict the light ray at $\Delta =0$. The same results about the
deflection of light ray have been discovered by us using the semiclassical
theory \cite{ZDL}.

From the point of particle nature, the force acting on the particle is
completely determined by the value and sign of $\eta _{1}$. Thus for a
particle passing through the point $a\neq 0$, when $\Delta =0$, the particle
does not feel any force, so it travels across straightly, so does the
particle at point $a=0$. For a particle traversing through the position $%
a\left( >0\right) $, when $\Delta <0$, this particle is subject to a
negative force, which moves the particle to the left with respect to
it original place; however, when $\Delta >0$, it experiences a
positive force, which makes the particle move to the right. Also for
a particle going through the place at $a(<0)$, when $\Delta <0$, the
particle moves to the right because of the action of a positive
force; when $\Delta >0 $, it goes to the left due to the action of a
negative force.

We have to point out that the magnetic field is not necessary for the
occurrence of the above-described phenomenon. For the model containing a $%
\Lambda $-type atomic ensemble interacting with one control beam and
one probe beam, similar phenomenon can also be found as long as the
two photon detuning
\end{subequations}
\begin{equation}
\Delta =\Delta _{p}-\Delta _{c}  \label{6-14}
\end{equation}%
varies, where $\Delta _{p}=\omega _{eg}-\nu $ is the detuning between the
atomic transition from $\left\vert e\right\rangle $ to $\left\vert
g\right\rangle $\ and the probe beam, $\Delta _{c}=\omega _{es}-\nu _{c}$ is
the detuning between the atomic transition from $\left\vert e\right\rangle $
to $\left\vert s\right\rangle $\ and the control beam. In order to clarify
the dependence of light deflection on two photon detuning $\Delta $, we
begin our description with the Hamiltonian in the interaction picture. In
the rotating frame with respect to
\begin{equation*}
H_{0}=\omega _{e}\sigma _{ee}+\left( \omega _{s}+\Delta _{c}\right) \sigma
_{ss}+\left( \omega _{g}+\Delta _{p}\right) \sigma _{gg}\text{,}
\end{equation*}%
the interaction Hamiltonian reads
\begin{equation*}
H_{I}^{\prime }=-\frac{N}{V}\int d^{3}r[\Delta _{p}\sigma _{gg}+\Delta
_{c}\sigma _{ss}+\left( g\sigma _{eg}E+\Omega \sigma _{es}+h.c.\right) ]%
\text{,}
\end{equation*}%
in the absence of magnetic field. The first order atomic transition
operators have the similar form as Eq.(\ref{3-09}) by replacing $\mu
_{s}B$, $\mu _{g}B$ and $\mu _{e}B$ by $\Delta _{c}$, $\Delta _{p}$
and zero respectively. Then the atomic transition operator $\sigma
_{ge}^{(1)}=g\Delta E/\left\vert \Omega \right\vert ^{2}$, which
induces a potential dependent on the two photon detuning $\Delta $.

In a real experiment, the dephasing rate of the forbidden $\left\vert
e\right\rangle $-$\left\vert s\right\rangle $ transition is nonzero due to
atomic collisions etc. Therefore an additional anti-Hermitian decay term
will be introduced phenomenologically into the effective Hamiltonian%
\begin{equation}
H_{eff}=cp_{z}+\frac{c}{2k}p_{x}^{2}+\eta _{0}^{\prime }+\eta _{1}^{\prime
}\left( x-a\right) \text{,}  \label{6-15}
\end{equation}%
where $\eta _{j}^{\prime }=-a_{j}-ib_{j}$, $j=0,1$ are complex. Then
it can be found that, after light passing through the Rb gas cell,
the dephasing rate introduces two additional terms to Eq.~(\ref{6-05}a)
\begin{equation}
x=a+\frac{a_{1}c}{2k}T^{2}-b^{2}b_{1}T-\frac{b_{1}c^{2}}{b^{2}k^{2}}T^{3}
\label{6-16}
\end{equation}
where $T=L/c$ is the time for the light travelling through the medium, and
\begin{eqnarray*}
a_{1} &=&\frac{2a}{\sigma ^{2}}\frac{N\left\vert g\right\vert ^{2}}{\Omega
_{0}^{2}}\exp \left( \frac{a^{2}}{\sigma ^{2}}\right) \Delta \text{,} \\
b_{1} &=&\frac{2a}{\sigma ^{2}}\frac{N\left\vert g\right\vert ^{2}}{\Omega
_{0}^{2}}\exp \left( \frac{a^{2}}{\sigma ^{2}}\right) \gamma _{2}\text{.}
\end{eqnarray*}%
For an atomic medium with length $L=7.5cm$ and density $N/V=10^{12}cm^{-3}$,
the dephasing rate $\gamma _{2}\approx 10^{-4}\gamma _{1}$. When a control
beam with width $\sigma =L/4$ and frequency $\nu _{c}=5\times 10^{14}Hz$ is
coupled to the atomic ensemble with $\Omega _{0}=5\gamma _{1}$, for the
probe beam with width $0.07mm$ incident at position $a=\sigma /2$, two
additional terms $b^{2}b_{1}T\approx 10^{-3}$ and $%
b_{1}c^{2}T^{3}/(b^{2}k^{2})\approx 10^{-5}$ caused by dephasing are much
smaller than that of the term $a_{1}cT^{2}/(2k)\approx 5\times 10^{-2}$
induced by the frequency detuning $\Delta $, which means Eq. (\ref{6-05}a)
dominates the distance in the transverse direction.

Next we investigate how the trajectory of the probe beam behaves when an
effective potential include the quadratic term of $x$ and $y$ in the
transverse direction. This induced potential is obtained when we expand
\begin{equation}
\left\vert \Omega \right\vert ^{-2}=\Omega _{0}^{-2}\exp \left[ \frac{x^{2}}{%
\sigma _{x}^{2}}+\frac{y^{2}}{\sigma _{y}^{2}}\right]   \label{6-06}
\end{equation}%
around the center $a_{x}$ and $a_{y}$ of the incident beam with the profile
shape
\begin{equation}
E\left( 0\right) =\frac{1}{\sqrt{\pi b^{2}}}\exp \left[ \frac{z^{2}+\left(
x-a_{x}\right) ^{2}+\left( y-a_{y}\right) ^{2}}{-2b^{2}}\right] \text{.}
\label{6-07}
\end{equation}%
By retaining the quadratic term of $x-a_{x}$ and $y-a_{y}$. The paraxial
motion equation becomes%
\begin{equation}
i\partial _{t}E+ic\partial _{z}E+\frac{c}{2k}\left( \partial
_{x}^{2}+\partial _{y}^{2}\right) E=\left[ V\left( x\right) +V\left(
y\right) \right] E  \label{6-08}
\end{equation}%
where
\begin{equation}
V\left( \chi \right) =\left[ \zeta _{\chi 0}+\zeta _{\chi 1}\left( \chi
-a_{\chi }\right) +\zeta _{\chi 2}\left( \chi -a_{\chi }\right) ^{2}\right] E%
\text{.}  \notag
\end{equation}%
The coefficients for $\chi =\left\{ x,y\right\} $ are
\begin{subequations}
\label{6-09}
\begin{align}
\zeta _{\chi 0}& =-\Omega _{0}^{-2}\exp \left[ \frac{a_{\chi }^{2}}{\sigma
_{\chi }^{2}}\right] \text{,} \\
\zeta _{\chi 1}& =-\frac{2|g|^{2}N}{\sigma _{\chi }^{2}}a_{\chi }\Delta
\Omega _{0}^{-2}\exp \left[ \frac{a_{\chi }^{2}}{\sigma _{\chi }^{2}}\right]
\text{,} \\
\zeta _{\chi 2}& =-\frac{\sigma _{\chi }^{2}+2a_{\chi }^{2}}{2\sigma _{\chi
}^{4}}\Omega _{0}^{-2}|g|^{2}N\Delta \exp \left[ \frac{a_{\chi }^{2}}{\sigma
_{\chi }^{2}}\right] \text{.}
\end{align}%
After a period of time, the Gaussian state will evolve into
\end{subequations}
\begin{equation}
E\left( t\right) =U\left( t\right) E\left( 0\right)
\end{equation}%
where the evolution operator
\begin{equation}
U\left( t\right) =\exp \left[ -i\left( cP_{z}+\frac{P_{x}^{2}+P_{y}^{2}}{%
2m^{\prime }}+V\left( x\right) +V\left( y\right) \right) t\right]
\end{equation}

Here, we assume that the detuning $\Delta$ is always negative. The Schr\"{o}
dinger-type equation (\ref{6-08}) governs the evolution of the wave-function
of the signal light in the atomic medium. And the trajectory of light ray is
described by the mean value of the coordinate operator
\begin{equation}
\chi_{c}=\left\langle \chi\right\rangle =\int E^{\ast}(t)\chi E(t)d\chi
\text{,}  \label{6-09a}
\end{equation}
with $\chi\in\left\{ x,y,z\right\} $. An explicit calculation gives (see Appendix B)
\begin{subequations}
\label{6-10}
\begin{align}
x_{c} & = a_{x}-\zeta_{x1}\frac{1-\cos\left( \omega_{ox}t\right) }{%
m^{\prime}\omega_{ox}^{2}}\text{,} \\
y_{c} & = a_{y}-\zeta_{y1}\frac{1-\cos\left( \omega_{oy}t\right) }{%
m^{\prime}\omega_{oy}^{2}}\text{,} \\
z_{c} & = ct
\end{align}
with the angular frequency
\end{subequations}
\begin{equation}
\omega_{ox}=\sqrt{2\zeta_{x2}/m^{\prime}}\text{, }\omega_{oy}=\sqrt {%
2\zeta_{y2}/m^{\prime}}\text{.}  \label{6-11}
\end{equation}

As the light travels across the atomic medium, the light ray - the center of
the wave packet oscillates around the initial center in the transverse
direction
\begin{subequations}
\begin{align}
x_{c}& =a_{x}+\frac{\zeta _{x1}}{m^{\prime }\omega _{ox}^{2}}\left( \cos
\frac{\omega _{ox}z_{c}}{c}-1\right) \text{,} \\
y_{c}& =a_{y}+\frac{\zeta _{y1}}{m^{\prime }\omega _{oy}^{2}}\left( \cos
\frac{\omega _{oy}z_{c}}{c}-1\right) \text{.}
\end{align}%
The anisotropic motion and potential in Eq.~(\ref{6-08}) result in that,
light travels in a straight line in the $z$-direction since it acts as an
\begin{figure}[tbp]
\includegraphics[width=6 cm]{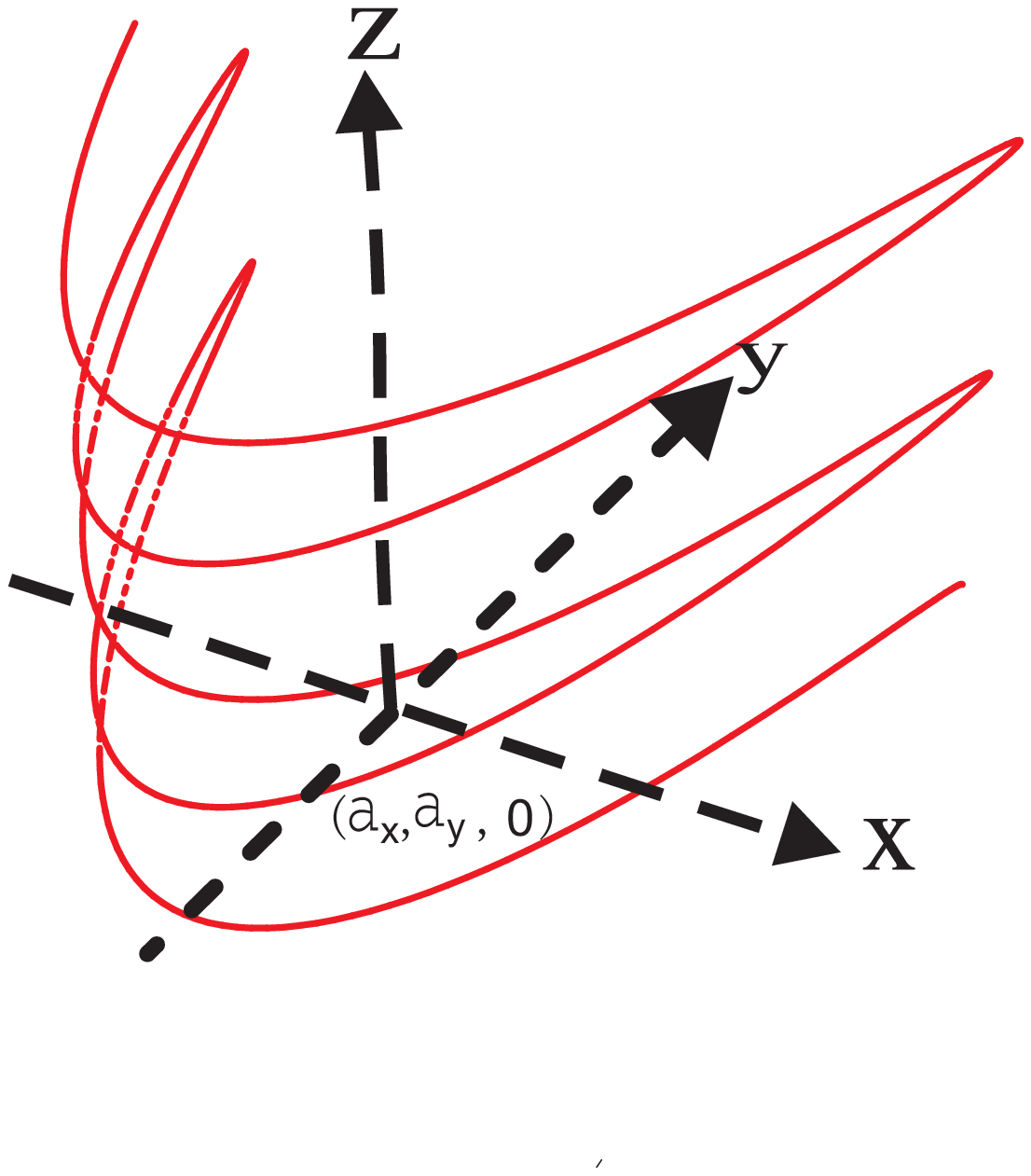}
\caption{\textit{(Color on line)}Schematic illustration about the ray
trajectory of the probe light in three dimension space.}
\label{fig:4}
\end{figure}
ultra-relativistic particle with velocity $c$, however the light oscillates
in the $x$-$y$ plane because it behaves as a non-relativistic particle with
effective transverse mass $m^{\prime }=k/c$. If $\zeta _{x2}=\zeta _{y2}$,
the light ray is a line with finite length in the transverse direction. In
Fig. \ref{fig:4}, we schematically illustrate the wave packet center of the
probe light in three-dimension space when $\zeta _{x2}\neq \zeta _{y2}$.

\section{\label{sec:sum}Summary}

In conclusion, we have developed a quantum approach for the spatial behavior
of propagating light when it passes through an EIT system with
spatial-dependent external fields. By studying the dynamics of the
atomic ensemble and the light pulse, the effective Schr\"{o}dinger equation
is derived to depict the space-time evolution of quasi-particles where the
effective potential is induced through the steady atomic response in the
external spatial-dependent fields. For a magnetic field with a spatial
distribution in the transverse direction, by considering the evolution of
the Gaussian state, we showed that: 1) in a harmonic magnetic field, the
light trajectory is a straight line; 2) in a linear magnetic field, the
light ray bends to the direction where the magnetic gradient increases. And
the deflection angle depends on four external parameters: the mixing angle
between the signal field and the collective atomic polarization, the wave
number of the signal pulse, the length of the EIT gas cell, and the small
magnetic field gradient. In an inhomogeneous optical control field, we
predict some novel results accessible on the light ray behavior. In the
linear response limit, it is found that the deflection of the light ray can
be controlled by two controllable external parameters: the center position
of the probe beam with respect to the control light, and the two photon
detuning. In the quadric expansion of coupling amplitude, the light
trajectory generally oscillates in the atomic medium.

Finally we note that our study is based on a quantum theoretical approach.
In our previous paper \cite{ZDL}, the Fermat principle is applied to study
the light trajectory in this atomic medium, and similar results are
obtained, but that approach is a semi-classical theory , which can be
understood in terms of the eikonal equation with the optical WKB
approximation of our approach. Though this semi-classical approach can
explain the most recent experiment \cite{Karpa} about the light deflection
by an EIT-based rubidium gas, it can not be further developed for the
investigation of photon state quantum storage since the signal light was
assumed as a classical field. The quantum approach assumes the probe light
is quantized, thus this approach can be used to investigate the possibility
for realizing a protocol for quantum sate storage with
spatially-distinguishable channels based on the EIT-enhanced light
deflection.

This work is supported by the NSFC with Grants No. 90203018, No. 10474104,
No. 60433050 and No. 10704023, NFRPC with Grant No. 2001CB309310 and
2005CB724508. One (LZ) of the authors also acknowledges the support of K. C.
Wong Education Foundation, Hong Kong. We acknowledge the useful discussions
with P. Zhang, T. Shi, and H. Ian.

\appendix

\section{Factorization of unitary operator $U_{li}$}

Since the unitary operator $U_{li}$ is exponential, factorizing it means to
express the exponential of a sum of operators in terms of a product of the
exponentials of operators. The unitary operator $U_{li}$ is an element of
the group \ generated by the momentum operators $P_{z}$, $P_{x}$ and the
coordinate $x$. Since $P_{z}$ commutes with $P_{x}$ and $x$, the unitary
operator can first be factorized as
\end{subequations}
\begin{equation}
U_{li}=e^{-iv_{g}tP_{z}}U_{li}^{\prime },  \label{app-01}
\end{equation}%
where
\begin{equation}
U_{li}^{\prime }=\exp \left[ -i\left( \frac{1}{2m}P_{x}^{2}-\mu b_{0}-\mu
\zeta x\right) t\right]  \label{app-02}
\end{equation}%
only contains operator $P_{x}$ and $x$, which generate the Lie algebra with
the basis $\{x,P_{x},P_{x}^{2},\mathbf{1}\}.$ Thus, operator $U_{li}^{\prime
}$ can be factorized as the form
\begin{equation}
U_{li}^{\prime }=e^{g_{1}P_{x}^{2}}e^{g_{2}P_{x}}e^{g_{3}x}e^{g_{4}},
\label{app-03}
\end{equation}%
and $g_{i}=g_{i}(t)$ are unknown functions of time $t$ to be determined.

Mathematically, the above factorization Ansatz is based on the Wei-Norman
algebraic theorem \cite{swna}: if the Hamiltonian of a quantum system%
\begin{equation}
H=\sum_{j=1}^{K}C_{j}(t)X_{j}  \label{app-04}
\end{equation}%
is a linear combination of the operators $X_{j}$ that can generate a $N$%
-dimensional Lie algebra with the basis:
\begin{equation}
\{X_{1},X_{2},...,X_{k},X_{k-1},....X_{N}\},  \label{app-05}
\end{equation}%
then the evolution operator governed by $H$ can be factorized as a product
of the single parameter subgroups, that is ,
\begin{equation}
U=\prod\limits_{j-1}^{N}e^{\xi _{j}(t)X_{j}},  \label{app-06}
\end{equation}%
where the coefficients $\xi _{j}(t)$ can be determined the \textquotedblleft
external field parameters\textquotedblright\ $C_{j}(t)$ through a system of
non-linear equations.

Now, we differentiate (\ref{app-03}) with respect to $t$ and the multiply
the resulting expression on the right hand side by the inverse of (\ref%
{app-03}), obtaining
\begin{subequations}
\begin{align}
& \frac{1}{2m}P_{x}^{2}-\mu \zeta x-\mu b_{0}=  \label{app-07} \\
& i\frac{\partial g_{1}}{\partial t}P_{x}^{2}+iP_{x}\left( \frac{\partial
g_{2}}{\partial t}-2ig_{1}\frac{\partial g_{3}}{\partial t}\right)  \notag \\
& +ix\frac{\partial g_{3}}{\partial t}+i\left( \frac{\partial g_{4}}{%
\partial t}-ig_{2}\frac{\partial g_{3}}{\partial t}\right) .  \notag
\end{align}

This leads to a systems of coupled differential equations
\end{subequations}
\begin{subequations}
\label{app-08}
\begin{align}
& i\frac{\partial g_{1}}{\partial t}=\frac{1}{2m}\text{, } \\
& i\frac{\partial g_{3}}{\partial t}=\mu \zeta \text{, } \\
& i\left( \frac{\partial g_{2}}{\partial t}-2g_{1}i\frac{\partial g_{3}}{%
\partial t}\right) =0, \\
& i\left( \frac{\partial g_{4}}{\partial t}-i\frac{\partial g_{3}}{\partial t%
}g_{2}\right) =\mu b_{0}.
\end{align}

The solution to these equations reads
\end{subequations}
\begin{subequations}
\label{app-09}
\begin{align}
g_{1}& =-i\frac{t}{2m}\text{, } \\
g_{3}& =it\mu \zeta , \\
g_{2}& =i\frac{\mu \zeta }{2m}t^{2}\text{,} \\
g_{4}& =i\mu \left( b_{0}t-\frac{t^{3}}{3}\frac{\mu \zeta ^{2}}{2m}\right) .
\end{align}%
Therefore the unitary operator $U_{li}$ is factorized into the form given in
Eq. (\ref{5-04}).

\section{Calculation of light trajectory in quadratic potential}

In the presence of the quadratic term of coordinates, the evolution operator
$U\left( t\right) $ in Eq. (\ref{6-08}) is generated by the effective
Hamiltonian $H_{3}=H_{z}+H_{y}+H_{x}$:
\end{subequations}
\begin{align}
H_{z} & = cP_{z}\text{,} \\
H_{y} & = \frac{P_{y}^{2}}{2m^{\prime}}+V\left( y\right) \text{,} \\
H_{x} & = \frac{P_{x}^{2}}{2m^{\prime}}+V\left( x\right) \text{.}
\end{align}
The evolution operator can be factorized as
\begin{equation}
U\left( t\right) =U_{z}\left( t\right) U_{y}\left( t\right) U_{x}\left(
t\right)
\end{equation}
with
\begin{align}
U_{z} & = U_{z}\left( t\right) =e^{-iH_{z}t}\text{,} \\
U_{y} & = U_{y}\left( t\right) =e^{-iH_{y}t}\text{,} \\
U_{x} & = U_{x}\left( t\right) =e^{-iH_{x}t}\text{.}
\end{align}
The expectation value of the coordinator along $x$-direction is calculated as%
\begin{align}
\left\langle x\right\rangle & = \int E^{\ast}\left( 0\right) U^{\dag }\left(
t\right) xU\left( t\right) E\left( 0\right) dx  \label{ap2-03} \\
& = \int E^{\ast}\left( 0\right) U_{x}^{\dag}xU_{x}E\left( 0\right) dx
\notag \\
& = a_{x}+\int E^{\ast}\left( 0\right) U_{x}^{\dag}\left( x-a_{x}\right)
U_{x}E\left( 0\right) dx  \notag
\end{align}
where commutation relation $\left[ x,P_{\chi}\right] =i\delta_{\chi x}$ is
used to obtain the second identity in Eq. (\ref{ap2-03}).

Actually, the effective Hamiltonian $H_{x}$ describes a harmonic oscillator
with its origin displaced from $a_{x}$ to other place. Thus, we rewrite the
coordinate and momentum operators in terms of the creation and annihilation
operator $\left\{ d^{\dag},d\right\} $:
\begin{align}
\eta & = \frac{1}{\sqrt{2m^{\prime}\omega_{ox}}}\left( d^{\dag}+d\right) , \\
P_{\eta} & = i\sqrt{\frac{m^{\prime}\omega_{ox}}{2}}\left( d^{\dag }-d\right)
\end{align}
with the inverse relation%
\begin{align}
d & = \sqrt{\frac{m^{\prime}\omega_{ox}}{2}}\eta+i\sqrt{\frac{1}{2m^{\prime
}\omega_{ox}}}P_{\eta}, \\
d^{\dag} & = \sqrt{\frac{m^{\prime}\omega_{ox}}{2}}\eta-i\sqrt{\frac {1}{%
2m^{\prime}\omega_{ox}}}P_{\eta},
\end{align}
where $\eta=x-a_{x}$ and $\omega_{ox}$ is given in Eq. (\ref{6-11}). The
effective Hamiltonian $H_{x}$ can be diagonalized as
\begin{equation}
H_{x}=\omega_{ox}\left( d^{\dagger}d+\frac{1}{2}\right)
\end{equation}
by a displacement operator
\begin{equation}
D\left( \beta\right) =\exp\left( \beta d^{\dagger}-\beta^{\ast}d\right)
\end{equation}
with%
\begin{equation}
\beta=\beta^{\ast}=-\frac{\zeta_{x1}}{\omega_{ox}\sqrt{2m^{\prime}\omega_{ox}%
}}\text{.}
\end{equation}
Then the evolution operator is factorized as the product of three operators
\begin{equation}
U_{x}=D\left( \beta\right) e^{-i\omega_{ox}\left( d^{\dagger}d+\frac{1}{2}%
\right) t}D^{-1}\left( \beta\right) \text{.}
\end{equation}
In terms of the creation and annihilation operators, the center of the light
wave packet
\begin{align}
\left\langle x\right\rangle & = a_{x}+\frac{\cos\left( \omega_{ox}t\right) }{%
\sqrt{2m^{\prime}\omega_{ox}}}\int E^{\ast}\left( 0\right) \left(
d^{\dagger}+d\right) E\left( 0\right) dx  \notag \\
& +i\frac{\sin\left( \omega_{ox}t\right) }{\sqrt{2m^{\prime}\omega_{ox}}}%
\int E^{\ast}\left( 0\right) \left( d^{\dagger}-d\right) E\left( 0\right) dx
\notag \\
& -2\frac{\beta\cos\left( \omega_{ox}t\right) }{\sqrt{2m^{\prime}\omega_{ox}}%
}+\frac{2\beta}{\sqrt{2m^{\prime}\omega_{ox}}}
\end{align}
Back to the coordinate and momentum operators, the center of the wave packet
in $x$-direction is obtained as Eq. (\ref{6-10}a). In a similar way, Eq. (%
\ref{6-10}b) can be achieved.

\end{document}